\documentclass{emulateapj}

\usepackage{color} 
\usepackage{graphics,graphicx}
\slugcomment{Accepted for publication in ApJ on March 9, 2016}
\shorttitle{Extreme Starburst Outflows}
\shortauthors{Heckman et al.}

\begin{document}

\title{The Implications of Extreme Outflows from Extreme Starbursts}
\author{Timothy M. Heckman \& Sanchayeeta Borthakur}
\affil{Center for Astrophysical Sciences, Department of Physics \& Astronomy, Johns Hopkins University, Baltimore, MD 21218}

\begin{abstract}

Interstellar ultraviolet absorption-lines provide crucial information about the properties of galactic outflows. In this paper, we augment our previous analysis of the systematic properties of starburst-driven galactic outflows by expanding our sample to include a rare population of starbursts with exceptionally high outflow velocities. In principle, these could be a qualitatively different phenomenon from more typical outflows. However, we find that instead these starbursts lie on, or along the extrapolation of, the trends defined by the more typical systems studied previously by us. We exploit the wide dynamic range provided by this new sample to determine scaling relations of outflow velocity with galaxy stellar mass (M$_*$), circular velocity, star-formation rate (SFR), SFR/M$_*$, and SFR/area. We argue that these results can be accommodated within the general interpretational framework we previously advocated, in which a population of ambient interstellar or circum-galactic clouds is accelerated by the combined forces of gravity and the momentum flux from the starburst. We show that this simple physical picture is consistent with both the strong cosmological evolution of galactic outflows in typical star-forming galaxies and the paucity of such galaxies with spectra showing inflows. We also present simple parameterizations of these results that can be implemented in theoretical models and numerical simulations of galaxy evolution.

\end{abstract}


\section{Introduction}

Galactic outflows driven by the energy and momentum supplied by a population of short-lived massive stars play crucial roles in the evolution of galaxies and the inter-galactic medium (Somerville \& Dav$\acute{e}$ 2015 and references therein). They can provide feedback that expels existing gas-phase baryons from galaxies and can prevent or inhibit the accretion of new gas from the circum-galactic or inter-galactic medium. This feedback can help explain the small ratio of baryons to dark matter observed in low-mass galaxies (e.g. McGaugh et al. 2010). In addition, the selective loss of newly-synthesized heavy elements from shallow potential wells is a key factor in the galaxy mass-metallicity relation (e.g. Tremonti et al. 2004; Andrews \& Martini 2013). The outward transport of low-angular momentum gas in outflows may help govern the relationship between the mass and size of galactic disks (e.g. Kauffmann et al. 2003). Galactic outflows have also been effective at transporting metals and even dust out into the circum-galactic and inter-galactic medium (e.g. M$\acute{e}$nard et al. 2010).

While there is general qualitative agreement about the importance of galactic outflows, it has not been possible so far to reliably quantify their effects. Part of the problem is that numerical simulations of galaxy evolution in a cosmological context cannot incorporate the physics of galactic outflows in a fully {\it ab initio} manner. Instead, both simulations and semi-analytic models are generally forced to rely on simple parameterizations of the relevant processes (Somerville \& Dav$\acute{e}$ 2015). Ideally, these would be based on a robust empirical characterization and theoretical understanding of galactic outflows, but neither of these yet exist.

One of the chief tools for investigating galactic outflows is spectroscopy using resonance lines in absorption to probe the cool and warm phases of the outflow (e.g. Heckman et al. 2000; Steidel et al. 2010).  This has the advantage that it can be readily applied to star-forming galaxies at both low-redshift (where our observational characterization of outflows is most complete) and high-redshifts (where galactic outflows are ubiquitous and strong).

Motivated by these considerations, we (Heckman et al. 2015 $–$ hereafter H15) recently analyzed high-quality ultraviolet spectroscopic data for a sample of 39 low-redshift starburst galaxies that spanned broad ranges in the principal properties of the galaxies and their starbursts. We found that some (and some combination) of these properties had strong systematic correlations with the observed properties of the outflows. We also showed that a simple model in which the outflowing gas seen in absorption is produced by a population of clouds accelerated by a combination of gravity and the momentum flux supplied by the starburst provided a good fit to the data.

In this paper, we seek to test these correlations and our model by extending our analysis to a population of 'extreme starbursts' that significantly extends the range in outflow velocity compared to the sample in H15. In principle, these high outflow velocities might require a qualitatively different model than that advanced in H15. For example these could be outflows driven by a powerful AGN (e.g. Liu et al. 2013) that has recently shut-down (e.g. LaMassa et al. 2015) or thermal instabilities in a radiatively-cooling fast outflow of themalized supernova and stellar wind ejecta (Thompson et al. 2016). 

We describe the new expanded sample in section 2 below, we highlight the empirical correlations in section 3, and briefly comment on the implications of these results in section 4.

\begin{deluxetable*}{cccccccc}

\tabletypesize{\scriptsize}
\tablecaption{Extreme Starburst Properties. \label{tbl-galaxy}}
\tablewidth{0pt}
\tablehead{
\colhead{Galaxy} & \colhead{$\rm Log~ M_*$} & \colhead{$\rm v_{cir}$} & \colhead{\rm $r_*$} & \colhead{$\rm  SFR$} & \colhead{$\rm Log~ SFR/M_*$} & \colhead{$\rm Log~ SFR/area$}  & \colhead{$\rm v_{max}$} \\
\colhead{}  & \colhead{($\rm Log~M_{\odot}$)} & \colhead{($\rm km~s^{-1}$)} & \colhead{($\rm pc$)} & \colhead{$(\rm M_{\odot} yr^{-1}$)}  & \colhead{$(\rm Log~ yr^{-1}$)} & \colhead{($\rm M_{\odot} yr^{-1} kpc^{-2}$)} & \colhead{($\rm km~s^{-1}$)}
}
\startdata
J0826+43 & 10.8 & 219 & 214 & 380 & -8.2 & 3.12 & 1230\\
J0905+57 & 10.7 & 204 & 91 & 260 & -8.3 & 3.70 & 2450\\
J0944+09 & 10.5 & 182 & 133 & 220 & -8.2 & 3.30 & 1330\\
J1104+59 & 10.6 & 191 & 219 & 70  & -8.8 & 2.37 & 1040\\
J1506+54 & 10.7 & 204 & 165 & 250 & -8.3 & 3.16 & 1480\\
J1506+61 & 10.2 & 148 & 217 & 210 & -7.9 & 2.85 & 1000\\
J1558+39 & 10.6 & 191 & 827 & 610 & -7.8 & 2.15 & 1000\\
J1613+28 & 11.2 & 288 & 980 & 230 & -8.8 & 1.58 & 1520\\
J1713+28 & 10.8 & 219 & 173 & 500 & -8.1 & 3.42 & 930
\enddata
\tablenotetext{a}{The galaxy circular velocity is derived from the stellar mass using the empirical relationship $\rm log~v_{cir} = 0.29~log~M_* - 0.79$, where $\rm v_{cir}$ is in km sec$^{-1}$, and $M_*$ is in solar masses. See text for details. All other parameters are taken from Sell et al. (2014) and Geach et al. (2014).}
\end{deluxetable*}

\section{Sample Properties}

In H15 we investigated a sample of 39 low-redshift ($z < 0.2$) starburst galaxies using observations of their ultraviolet interstellar absorption-lines made with the Far Ultraviolet Spectroscopic Explore (Grimes et al. 2009) and with the Cosmic Origins Spectrograph on the Hubble Space Telescope (Alexandroff et al. 2015).  These papers contain full descriptions of both the sample selection and the data analysis, and we refer the reader to them for details. In brief, about 60\% of the sample consisted 'Lyman Break Analogs' which have properties very similar to those of typical Lyman Break Galaxies at redshifts $z \sim$ 3 to 4 (e.g. Hoopes et al. 2007; Overzier et al. 2010).  The other 40\% of the sample consisted of ultraviolet-bright low-redshift galaxies representative of the local starburst population.

In this paper, we add a new sample of 'extreme starbursts' drawn from the recent investigations by Diamond-Stanic et al. (2012) and Sell et al. (2014; hereafter S14). These are intermediate redshift ($z \sim 0.4$ to 0.7) galaxies discovered in the Sloan Digital Sky Survey archive, characterized by very high outflow velocities (over $\sim 10^3$~km~s$^{-1}$), high star-formation rates (a few hundred M$\rm _{\odot}~yr^{-1}$) and compact sizes (starburst half-light radii of a few hundred pc).  These values lie well beyond beyond the ranges covered by the H15 sample. 

Based on their multi-waveband analysis, S14 concluded that the majority of these objects are energetically-dominated by the intense and compact starburst, with little or no contribution by an Active Galactic Nucleus (AGN). From their full sample of 12 galaxies, we have excluded the four that show any evidence for an AGN. We have also added the very similar galaxy investigated by Geach et al. (2014 – hereafter G14), giving us a sample of nine extreme starbursts.
 
While there have been a number of other investigations of outflows driven from intermediate redshift star-forming galaxies (e.g. Weiner et al. 2009; Erb et al. 2012; Kornei et al. 2012; Martin et al. 2012; Rubin et al. 2014; Bordoloi et al. 2014), these galaxies have been representative of typical star-forming galaxies at these epochs. Thus, the extreme-starburst sample complements these earlier studies by extending the investigation of outflows into an entirely new part of parameter space.

The principal properties of the new sample of extreme starbursts are listed in Table 1. The values for galaxy stellar mass (M$_*$), star-formation rate (SFR), starburst half-light radius ($r_*$), and maximum outflow velocity ($v_{max}$) are taken directly from S14 and G14. For the most part, these parameters were derived in a way that is consistent with the approach taken for the sample analyzed in H15. 

One exception is the definition of the outflow velocity. These were based on the Mg II 2796,2803 doublet for the extreme starbursts, and in many cases the doublet shows a significant amount of line-emission near, and redward of, the galaxy systemic velocity. In this case, in-filling of the absorption-line by emission can have a serious effect on the net profile (e.g. Prochaska et al. 2011; Martin et al. 2012; Erb et al. 2012; Kornei et al. 2012; Rubin et al. 2014; Scarlata \& Panagia 2015). This will mean that the flux-weighted centroid of the absorption lines will tend to over-estimate the outflow velocity. The maximum blueward extent of the absorption-line will not be affected by infilling, and so we adopt these values from S14. We have reanalyzed the COS and FUSE spectra described in H15 and measured the average value of the maximum outflow velocity based on the Si II 1260 and C II 1334 lines (COS) and the CII 1036 line (FUSE). These transitions arise from species that roughly match the ionization state of the Mg II ion. We will refer to these outflows velocities as v$_{max}$ to distinguish them from the flux-weighted mean outflow velocities ($v_{out}$) investigated in H15. We have estimated the uncertainty in v$_{max}$ by comparing the values measured individually for Si II 1190, Si III 1206, Si II 1260, and CII 1334 (COS data) and for C III 977, C II 1036, and N II 1084 (FUSE data). These values are listed in Table 2.

As in H15, we define the $\rm SFR/area$ to be 0.5 $SFR/\pi r_*^2$ and estimate the galaxy circular velocity ($v_{cir}$) based on the tight empirical relationship shown by the data presented by Simons et al. (2015) and then parameterized by H15: $\rm log~v_{cir} = 0.29~log~M_* - 0.79$, where $\rm v_{cir}$ is in km sec$^{-1}$, and $M_*$ is in solar masses.

\begin{figure*}

\includegraphics[trim=0mm 0mm 33mm 0mm,  clip=true,scale=0.6]{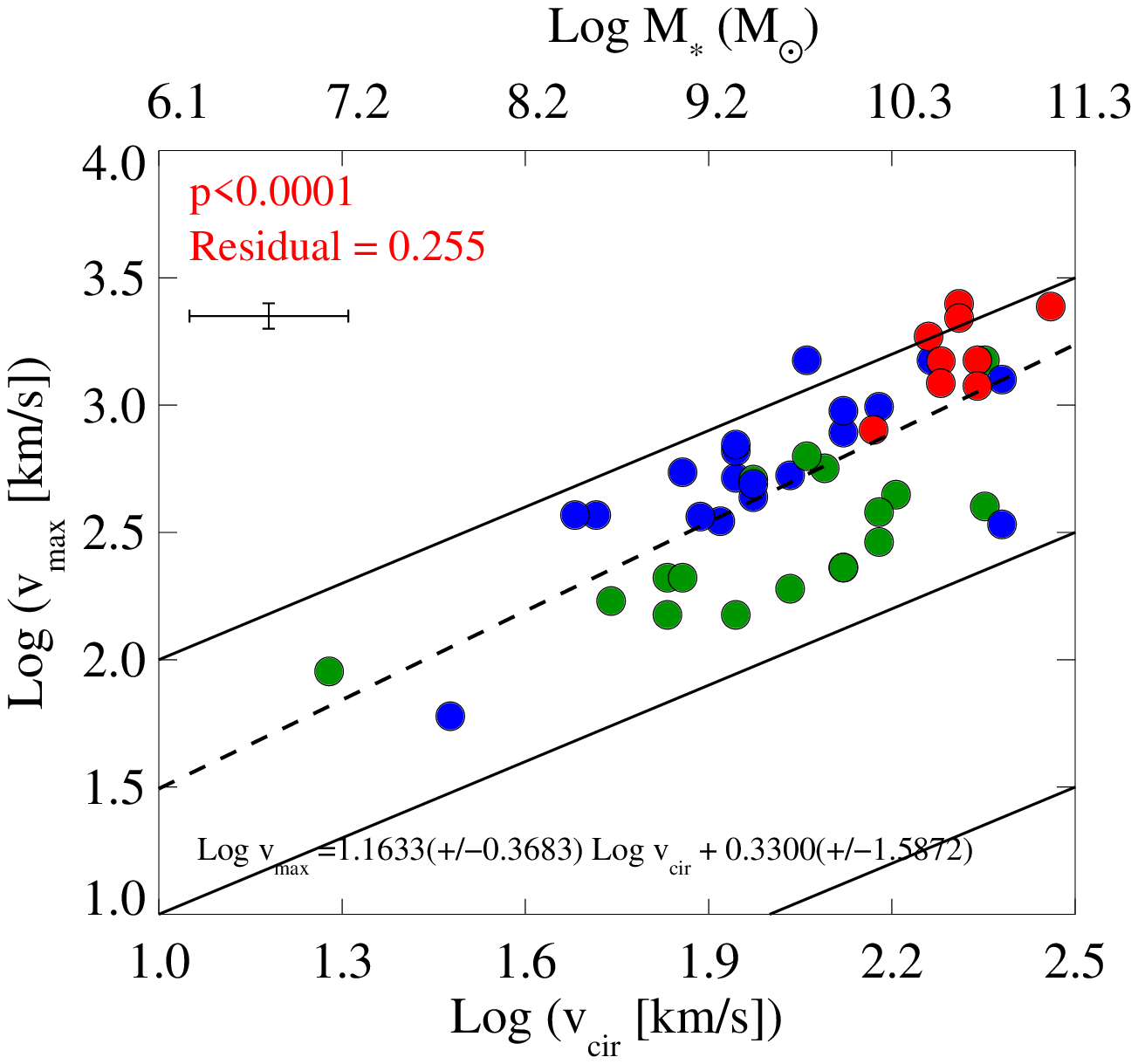}
\includegraphics[trim=0mm 0mm 33mm 0mm,  clip=true,scale=0.6]{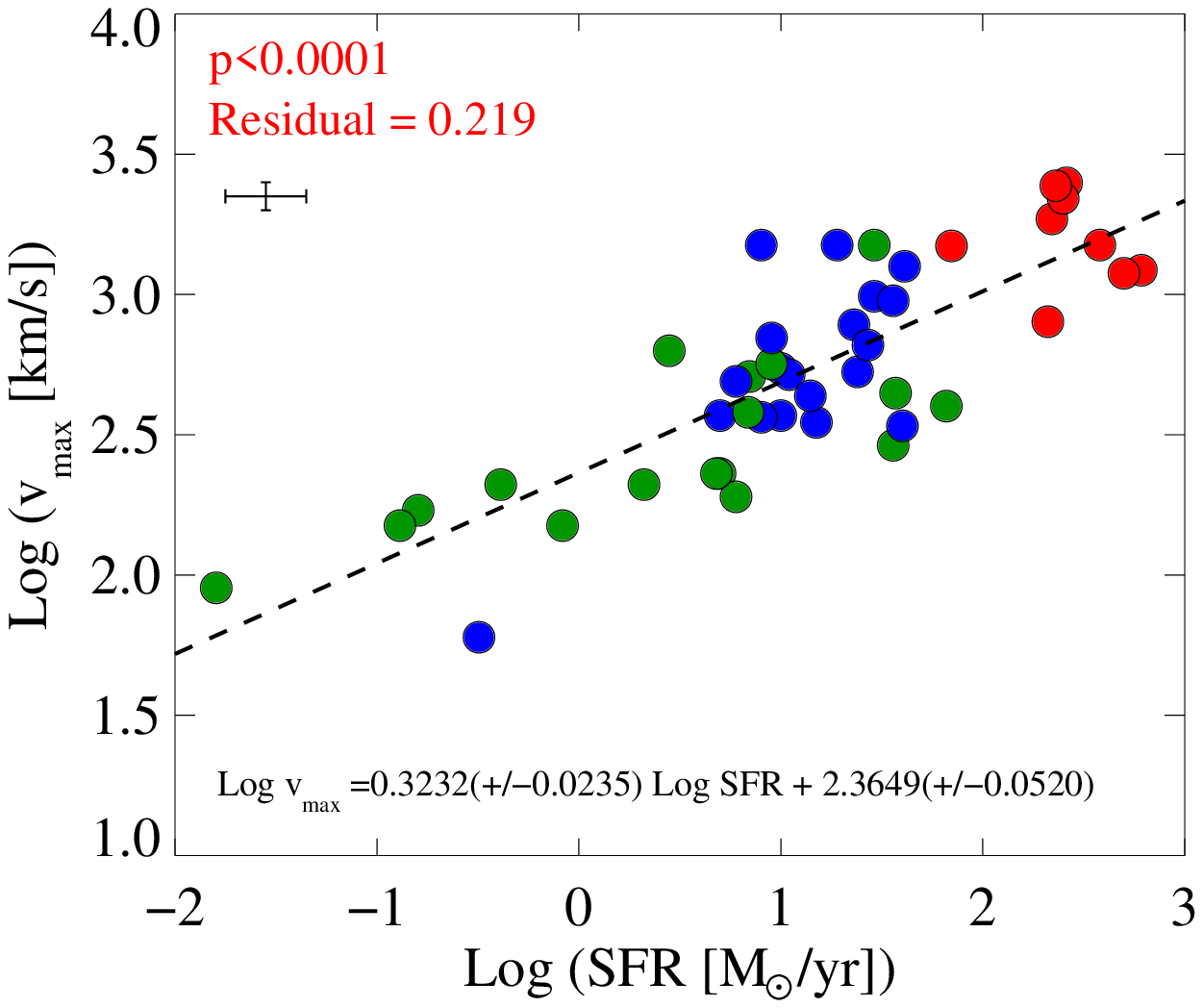}\\
\includegraphics[trim=0mm 0mm 33mm 0mm,  clip=true,scale=0.6]{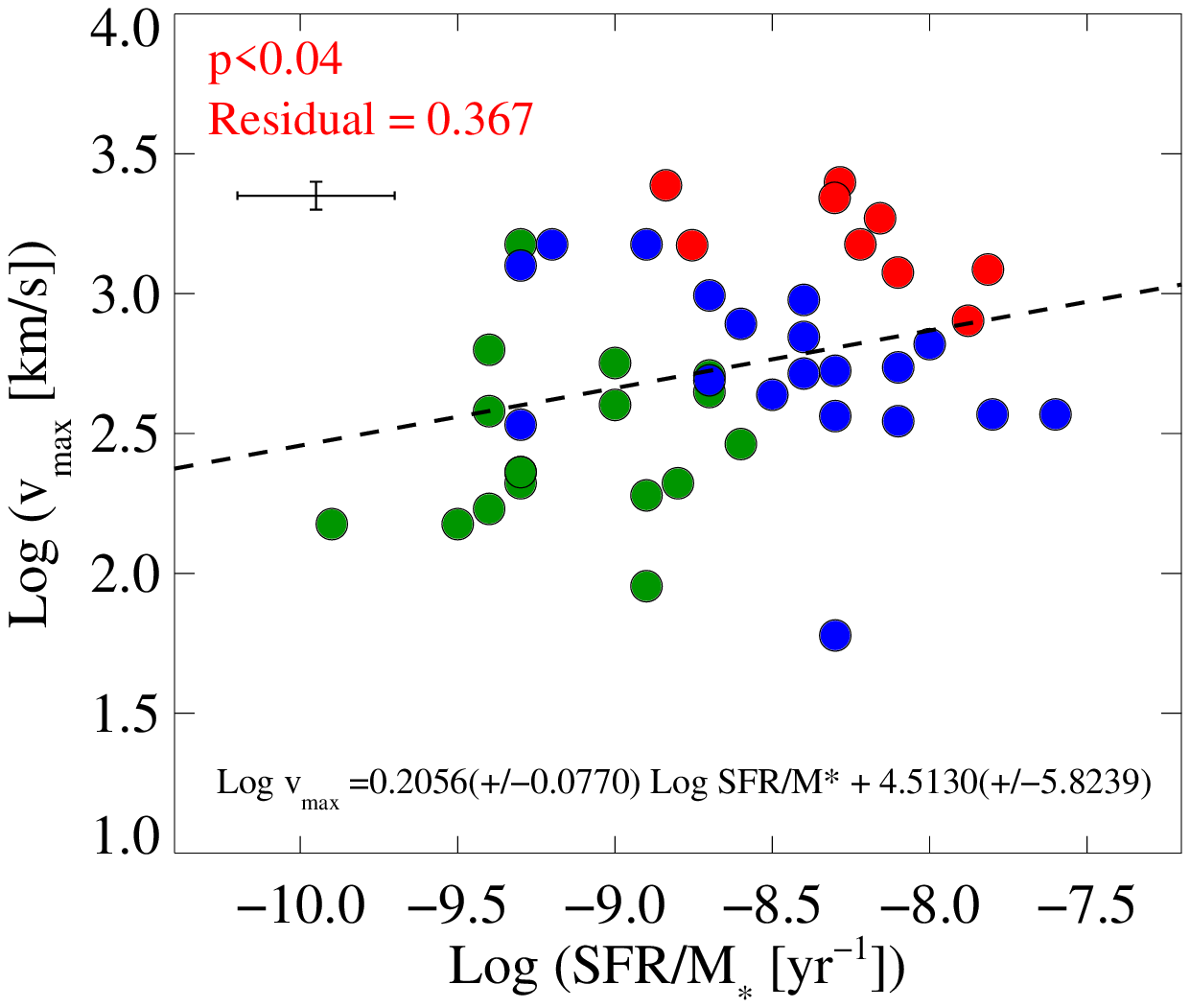}
\includegraphics[trim=0mm 0mm 33mm 0mm,  clip=true,scale=0.6]{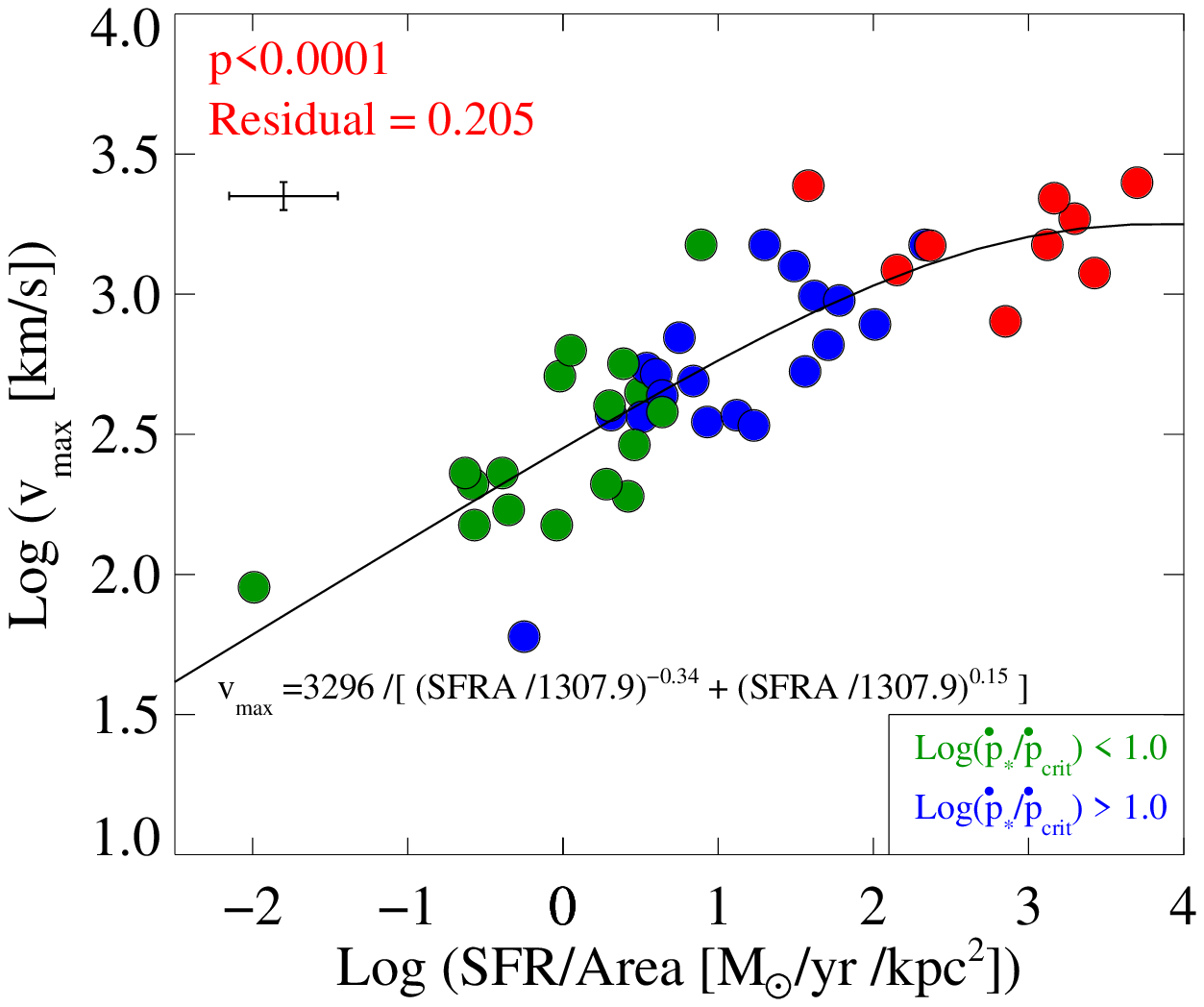}

\caption{ \scriptsize {The log of the maximum outflow velocity is plotted as a function of the basic properties of the starburst galaxies.  Clockwise from the upper left (a,b,c,d). The upper left panel (a) shows that there is a correlation between the maximum outflow velocity and the galaxy circular velocity. The diagonal lines show $v_{max} = 10~ v_{cir }$, $v_{max} = v_{cir}$, $v_{max} = 0.1 v_{cir}$. The label on the upper axis shows the corresponding values of the galaxy stellar mass (see text). The upper right panel (b) shows a strong correlation between SFR and $v_{max}$.  The bottom two panels show the correlation with two forms of normalized SFR: SFR/area (c) and SFR/M$_*$ (d). Both correlations are statistically significant, but the correlation with SFR/area is much stronger. The crosses represent the typical uncertainties (see H15 for details). The blue and green points show the strong- and weak-outflows from H15 and the red points show the extreme starbursts from S14 and G14. These lie along an extrapolation of the trends seen in H15 in all cases. In each panel we indicate the statistical significance of each correlation using the Kendall $\tau$ test. We also include the best-fit analytic function for each correlation (dashed lines) and the {\it rms} residuals in $log(v_{max})$ (data minus fit).}}
\end{figure*}

\begin{deluxetable}{lccccccc}
\tabletypesize{\scriptsize}
\tablecaption{Maximum Outflow Velocities. \label{tbl-oldgalaxy}}
\tablewidth{0pt}
\tablehead{
\colhead{Galaxy} &\colhead{$\rm v_{max}$} \\
\colhead{}  & \colhead{($\rm km~s^{-1}$)}
}
\startdata
J0021+00 & 350\\
J0055-00 & 530\\
J0150+13 & 450\\
J0213+12 & 1500\\
J0808+39 & 1500\\
J0823+28 & 370\\
J0921+45 & 1500\\
J0926+44 & 550\\
J0938+54 & 520\\
J1025+36 & 360\\
J1112+55 & 990\\
J1113+29 & 510\\
J1144+40 & 570\\
J1414+05 & 370\\
J1416+12 & 780\\
J1428+16 & 440\\
J1429+06 & 660\\
J1521+07 & 490\\
J1525+07 & 700\\
J1612+08 & 1000\\
J2103-07 & 1260\\
Haro~11  & 290\\
VV~114   & 400\\
NGC~1140 & 150\\
SBS~0335-052 & 60\\
Tol0440-381  & 230\\
NGC~1705  & 170\\
NGC~1741  & 190\\
I~Zw~18  &  90\\
NGC~3310  & 630\\
Haro~3  & 210\\
NGC~3690 & 340\\
NGC~4214 & 150\\
IRAS 19245+4140 & 210\\
NGC~7673 & 230\\
NGC~7714 & 380\\
\enddata
\end{deluxetable}

\section{Results}

\begin{figure*}
\includegraphics[trim=0mm 0mm 33mm 0mm,  clip=true,scale=0.6]{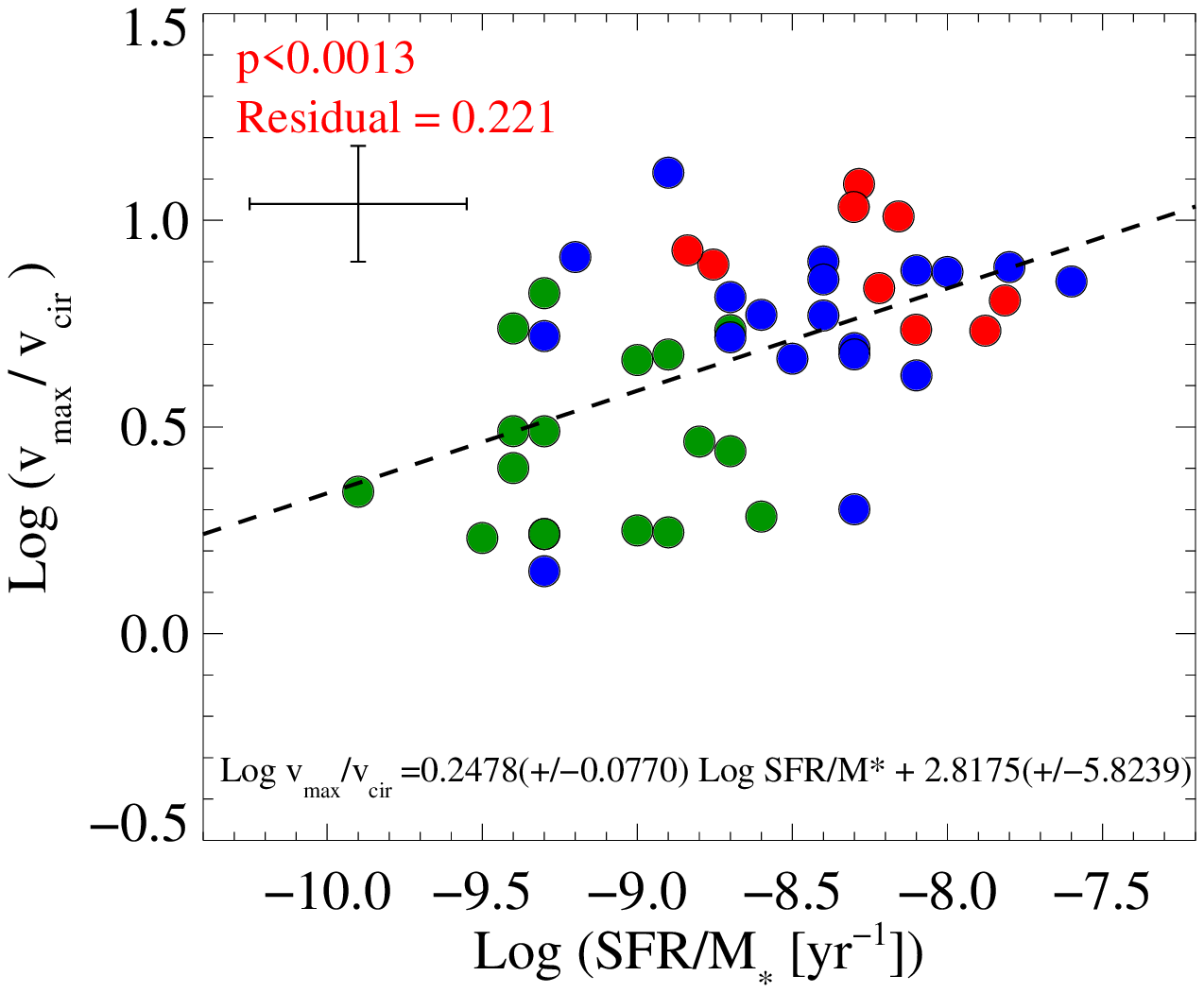} 
\includegraphics[trim=0mm 0mm 33mm 0mm,  clip=true,scale=0.6]{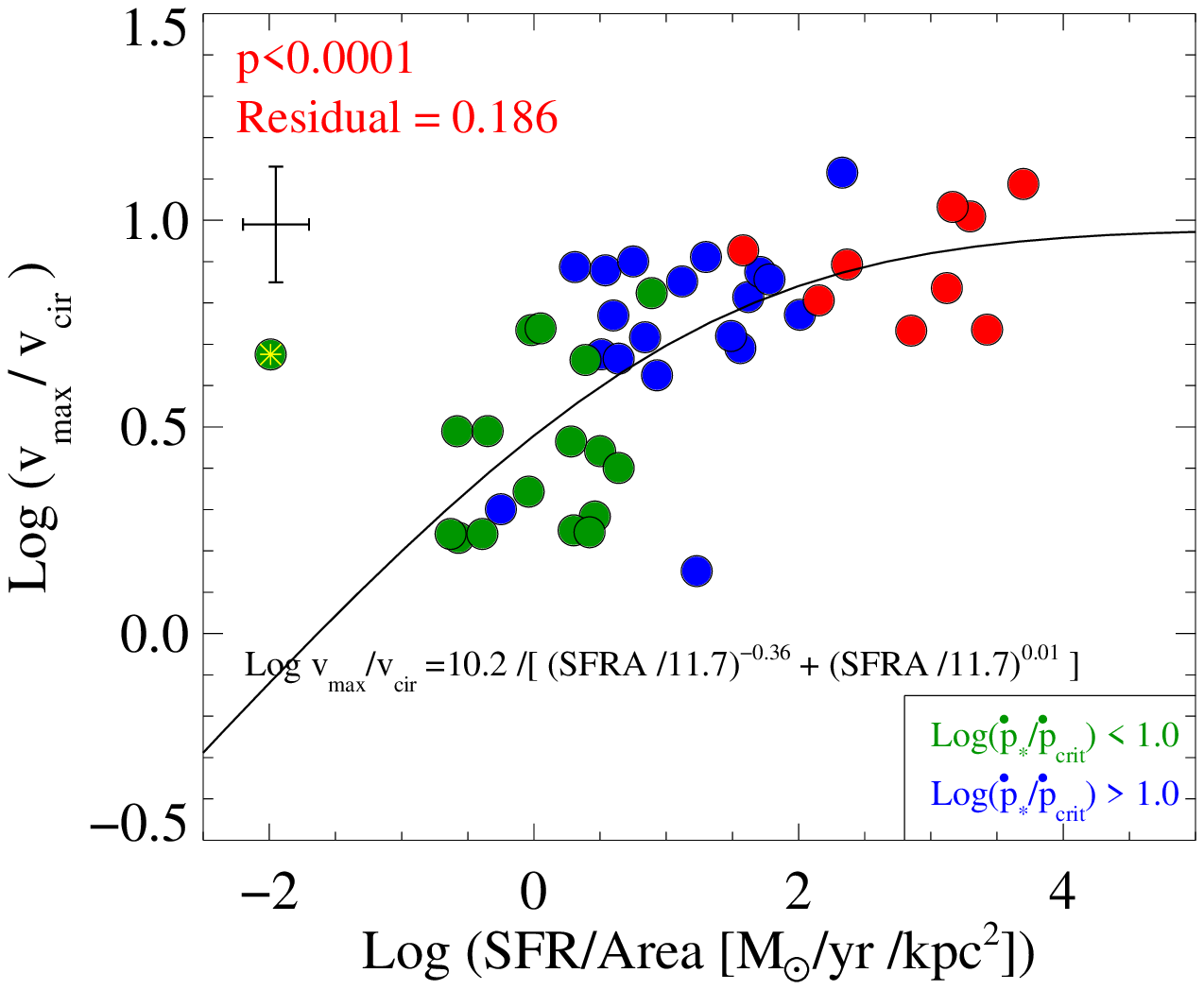}  
\caption{Right:  a) The maximum outflow velocity normalized by the galaxy circular velocity plotted as a function of SFR/area. Left: b)  $v_{max}/v_{cir}$  plotted as a function of the specific SFR (SFR/M$_*$). While both plots show statistically significant correlations, the correlation with SFR/area is much stronger. The crosses represent the typical uncertainties (see H15 for details). The blue and green points show the strong- and weak-outflows from H15 and the red points show the extreme starbursts from S14 and G14. These lie along an extrapolation of the trends seen in H15. In each panel we indicate the statistical significance of each correlation using the Kendall $\tau$ test. We also include the best-fit analytic function for each correlation (dashed lines) and the {\it rms} residuals in $log (v_{max}/v_{cir})$ (data minus fit)}  
 \end{figure*}

The motivation of this paper is to extend the analysis of outflows into the regime of the extreme starbursts. In Figure 1 we plot a set of correlations between the outflow velocity and the principal properties of the galaxies and their starbursts. It is immediately clear that the extreme starburst sample allows us to probe hitherto unexplored parts of parameter space.

Different investigations in the past have not always agreed with one another in terms of the strengths of the correlations of outflow velocity with SFR (Heckman et al. 2000; Martin 2005; Rupke et al. 2005; Martin et al. 2012; Weiner et al. 2009; Erb et al. 2012; Kornei et al. 2012; Rubin et al. 2014; Bordoloi et al. 2014; Chisholm et al. 2015; H15), with SFR/area (Chen et al. 2010; Kornei et al. 2012; Rubin et al. 2014; Chisholm et al. 2015; H15), and with either M$_*$ or v$_{cir}$ (Heckman et al. 2000; Rupke et al. 2005; Martin 2005; Erb et al. 2012; Martin et al. 2012; Rubin et al. 2014; Bordoloi et al. 2014; Chisholm et al. 2015; H15). 

The new sample allows us to reexamine the relations between outflow velocity and these properties over an unprecedented dynamic range. These relations are plotted in Figure 1. The first important conclusion is that in all four panels the extreme starbursts lie along an extrapolation of the relations defined by the more typical starbursts in the H15 sample. This provides indirect evidence that these extreme outflows are not a qualitatively different physical phenomenon.

Figure 1 shows that the maximum outflow velocities correlate most strongly with SFR/area, and least strongly with SFR/M$_*$. We can quantify this by determining simple analytic fitting relations to the correlations in the four panels. The relations with $v_{cir}~\rm (M_*$), SFR, and SFR/M$_*$ can all be fit as single power-law relation (which we give in each figure panel). The correlation with SFR/area 'saturates' at the high end, and so we fit this as a double powerlaw. We also list the {\it rms} residuals about the fits in each panel.    

The relationship between $v_{max}$ and $v_{cir}$ is particularly important in the context of models of galaxy evolution (Somerville \& Dav$\acute{'e}$ 2015 and references therein). The ratio of these velocities in Figure 1 varies by about an order-of-magnitude. We now show in Figure 2 that this ratio correlates strongly and systematically with the star formation rate (SFR/area), and less so with the specific SFR (SFR/M$_*$). The results are consistent with those presented in H15, but the relationships can now be probed over an increased dynamic range (particularly for SFR/area). The correlation with SFR/area also shows a saturation in normalized outflow velocity at $v_{max} \sim$ 6 to 10 $v_{cir}$ above SFR/area $\sim \rm 10^2~ M_{\odot}~yr^{-1}~kpc^{-2}$ . It is important to emphasize that outflow velocities this high are well in excess of the galaxy escape velocities. As in the case of Figure 1, we show the analytic fits to the relations and list the {\it rms} residuals about these fits.

\section{Discussion}

\subsection{The Model of Momentum-Driven Clouds}

We have shown that the extreme starbursts lie along the extrapolation of the trends defined by the more typical starbursts (Figures 1 and 2). We now want to discuss whether these results can be easily understood within the context of the basic physical picture described in H15.

H15 showed that a simple model of a population of clouds that are accelerated by the combination of gravity and the starburst momentum flux  provided a good description of the properties of galactic outflows. In particular, we defined a critical momentum flux such that the net force on a cloud located at $r = r_*$ (the launch point of the outflow) is outward:

\begin{equation}
\dot{p}_{crit} = \Omega r_* N_{c}  <m> v_{cir}^2
\end{equation}

Here, $\Omega$ is the solid angle occupied by the wind, $N_c$ is the cloud Hydrogen column density and $<m>$ is the mean mass per H atom. In convenient units $\dot{p}_{crit} = 10^{33.9}$ dynes for $\Omega = 4 \pi$, $N_c = 10^{21}~cm^{-2}$, $r_* = 1$ kpc, and $v_{cir} = 100$ km s$^{-1}$. 

The momentum flux from the starburst is a combination of contributions from both radiation pressure (e.g. Murray et al. 2005) and the ram pressure of a hot outflowing wind fluid collectively created from the ejecta of massive stars (Chevalier \& Clegg 1985). For a standard Kroupa/Chabrier initial mass function and a constant star-formation rate

\begin{equation}
\dot{p}_* = 4.8 \times 10^{33} SFR~ dynes
\end{equation}

where SFR is in M$_{\odot}$/year (H15). By expressing the SFR in gm s$^{-1}$ (hereafter, sfr), we can rewrite this as 

\begin{equation}
\dot{p}_* = v_{eff} sfr
\end{equation}

where $v_{eff} = \rm 7.6 \times 10^7~cm~s^{-1}$ (760 km/sec). 

We showed in H15 that the properties of the outflows depended strongly on the ratio of the momentum flux supplied by the starburst ($\dot{p}_*$) relative to the critical value($\dot{p}_{crit}$):  

\begin{equation}
R_{crit} = \dot{p}_*/\dot{p}_{crit} = sfr~v_{eff}/(\Omega r_* N_c <m> v_{cir}^2).
\end{equation}

In convenient units, $R_{crit} =$ 0.57 for SFR = 1 M$_{\odot}$ year$^{-1}$, $\Omega = 4 \pi$, $r_* =$ 1 kpc, $N_c = 10^{21}$ cm$^{-2}$, and $v_{cir} =$ 100 km sec$^{-1}$. We cannot directly compute $R_{crit}$ for the extreme starbursts, since have no estimate of $N_c$ for them. Nonetheless, the very high SFR and small sizes of these objects imply values for $R_{crit}$ at least as high of the upper end of the H15 sample ($R_{crit} > 10$). Indeed, in Fig. 1 and 2 the extreme outflows (red points) overlap much better with the H15 strong outflows (blue points, defined as $R_{crit} > 10$) than with the weak outflows (green points, $1 < R_{crit} < 10$). 

H15 derived the equation-of-motion for the clouds in the model we are considering here. In particular, they showed that the maximum outflow velocity for a cloud in an isothermal potential (which will occur at the radius at which the net radial force on the cloud is zero) could be written as:

\begin{equation}
v_{max,c}/v_{cir} = \sqrt{2} [(R_{crit}-1)-ln(R_{crit})]^{1/2}
\end{equation}

H15 showed that the predicted outflow velocities were a good match to the data. Without direct measurements of $R_{crit}$ for the extreme starbursts, we cannot extend this comparison to the extreme starbursts. However, as discussed in H15 equation 5) above predicts a rapid rise in outflow velocity for small values ($R_{crit} \sim 1$ to 10) and then a flattening in the relationship for $R_{crit} > 10$. Given that $R_{crit} \propto SFR/r_*$, this is almost certainly the underlying physics seen in the flattening of the upper end of the relationship between $v_{max}/v_{cir}$ and SFR/area (Figure 2).

We conclude that the very large outflow velocities seen in the extreme starbursts are at least qualitatively consistent with model of momemtum-driven clouds in H15 (taken into an extreme regime).

\subsection{Implications}

As we emphasized in section 1, having a secure empirical characterization and physical understanding of galactic outflows is an important step in being able to develop a quantitative assessment of their impact. We therefore close the paper by considering the implications of our results for understanding galaxy evolution.

First, it is instructive to recast $R_{crit}$ as defined in H15 and summarized above. We define the {\it dynamical} mass of the starburst as

\begin{equation}
 M_{sb} = 2 v_{cir}^2 r_*/G
\end{equation}

where this assumes that half the mass is enclosed within the half-light radius ($r_*$). This together with Equation 5 above then allows us to write:

\begin{equation}
R_{crit} = 2 sfr~v_{eff}/ \Omega G M_{sb} N_c <m>
\end{equation}

This equation shows that $R_{crit} \propto SFR/M_{sb} = sSFR_{sb}$. We emphasize that this is a specific star-formation rate {\it within the starburst} and is normalized with respect to a {\it dynamical mass}. In both respects, this is different from the specific star-formation rate pertaining to the {\it entire galaxy} and normalized to the {\it total galaxy stellar mass} (as plotted in Figures 1 and 2). In convenient units, $R_{crit} = 2.7 (4 \pi/\Omega)(sSFR_{sb}/$Gyr$^{-1}) (N_c/10^{21}$ cm$^{-2}$).
 
Low-redshift starbursts are usually compact and circum-nuclear (i.e. $M_{sb} << M_*$). This is not typically the case at high-redshift (e.g. Shapley 2011; Forster-Schreiber et al. 2011; Somerville \& Dav$\acute{e}$ 2015). If we therefore make the assumption that the galaxy-wide specific star-formation rate can be used to estimate $R_{crit}$ at high-redshift, the strong evolution in the specific star-formation rate with redshift (Madau \& Dickinson 2014 and references therein) implies a correspondingly large increase in $R_{crit}$ and hence in the importance of strong outflows. More quantitatively, adopting $N_c = 10^{21}$ cm$^{-2}$ would imply that the characteristic value of $R_{crit}$ for galaxies on the star-forming main sequence increases from $\sim$ 0.3 ($z \sim$ 0), to $\sim$ 3 ($z \sim 1$), to $\sim$ 5 ($z \sim 2$), and to $\sim$ 13 ($z \sim$ 4 to 7). This is at least qualitatively consistent with the rarity of strong outflows in the present-day universe and their near-ubiquity at $z >$ 2.

These results may also have implications for understanding why it has been difficult to find direct spectroscopic evidence for inflowing gas in strongly star-forming galaxies at intermediate and high redshift (e.g. Steidel et al. 2010; Martin et al. 2012; Rubin et al. 2012). To begin, we note that an infalling cloud must satisfy the condition that the inward force of gravity exceeds the outward momentum flux on the cloud from the starburst. This is just recasting Equation 7) above in terms of a critical column density for infall:

\begin{equation}
N_{inflow} > N_{crit} = 2 ssfr_{sb} v_{eff}/(\Omega G <m>)
\end{equation}

Here $ssfr_{sb}$ is in units of sec$^{-1}$ and $v_{eff}$ is 7.6 $\times 10^7$~cm~sec$^{-1}$. In convenient units this corresponds to $N_{crit} = 8 \times 10^{21}~cm^{-2}$ for $sSFR_{sb} = 10^{-9}$ year$^{-1}$ and $\Omega = 4 \pi$). Assuming a normal dust/metals ratio in this material (Mattson et al. 2014), the Calzetti et al. (2000) dust attenuation law would imply a far-UV extinction of $A_{fuv} = 25 Z/Z_{\odot}$ magnitudes for this column density. If the outflow flow is not spherically symmetric (e.g. it is bipolar or more generally follows the path-of-least resistance out of the galaxy the implied column density and dust extinction become even larger.

Thus, unless the gas has very sub-solar abundances (which might apply to relatively pristine gas being accreted from the cosmic web), it would be essentially opaque in the far-UV and therefore undetectable in the spectra. If this gas covered the whole far-UV source, the galaxy itself would be invisible in the far-UV and would not even enter a sample targeted for rest-frame far-UV spectroscopy in the first place.
  
\section{Conclusions}

The goal of this paper was to improve our understanding of galactic outflows by maximizing the dynamic range over which their properties can be probed. To that end, we have used observations of outflows traced by ultraviolet interstellar absorption-lines for a sample of nine extreme starbursts (Sells et al, 2014; Geach et al. 2014). More specifically, the extreme starbursts are characterized by significantly higher star-formation rates per unit area (SFR/area) than even the most extreme members of the sample of 39 low-redshift starbursts we studied previously (Heckman et al. 2015, hereafter H15). 

We found that in all respects, the extreme starbursts lay along a smooth extrapolation of the correlations seen in H15. The addition of the extreme starbursts strengthened the results in H15 by significantly expanding the dynamic range over which these correlations have been delineated.

Specifically, we found that the maximum outflow velocity ($v_{max}$) correlated most strongly with the SFR/area and least strongly with SFR/M$_*$. The ratio of $v_{max}/v_{cir}$ spanned about an order-of-magnitude and correlated strongly and positively with SFR/area (and less so with SFR/M$_*$). This ratio reached typical values of $\sim$ 3 to 10 for starbursts with high SFR/area, well in excess of the galaxy escape velocity. We exploited the large dynamic range spanned by our sample to derive simple analytic fits to all these empirical relations.

We then argued that the properties of the extreme starbursts were consistent with the simple analytic model for the outflows explored by H15 in which a population of clouds is accelerated by the net sum of gravity and the momentum-flux supplied by the starburst. H15 emphasized the importance of the ratio of the momentum flux supplied by the starburst to the minimum amout required to balance the inward force of gravity on a cloud ($R_{crit}$).

We showed that $R_{crit}$ is simply proportional to the value of SFR per unit dynamical mass evaluated over the star-forming region within the galaxy. We argued that the strong observed evolution in the specific star-formation rate with cosmic time then implies a strong evolution in the importance of strong outflows. This is at least qualitatively consistent with what we know observationally. We also showed that material meeting the criterion $R_{crit} < 1$ (which is required for infall) will have column densities of-order $10^{22}$ cm$^{-2}$ at intermediate and high redshift. This material would be opaque in the rest-frame far-UV, potentially explaining why the direct signature of infalling gas (redshifted absorption-lines) is only rarely detected.
  
\vspace{.5cm}

\acknowledgements 
TH thanks the Kavli Institute for Theoretical Physics, the Simons Foundation, and the Aspen Center for Physics for supporting informal workshops on galactic winds, feedback, and the circum-galactic medium during 2014 and 2015. These provided much of the motivation for this paper.  Conversations at these workshops with Romeel Dav{\'e}, Crystal Martin, Norm Murray, Ralph Pudritz, Eliot Quataert, Brant Robertson, Chuck Steidel, Todd Thompson, and Ellen Zweibel were especially helpful. He also thanks the Sackler Foundation, Churchill College, and the Institute of Astronomy for supporting his visit to the University of Cambridge. He thanks Martin Haehnelt, Max Pettini, and Mark Whittle at the IoA for especially useful conversations. Finally, we thank the referee for suggestions that helped us clarify the presentation and discussion of our results.

This work is based on observations with the NASA/ESA Hubble Space Telescope, which is operated by the Association of Universities for Research in Astronomy, Inc., under NASA contract NAS5-26555. The data analysis was supported by grant HST GO 12603. This work was supported in part by National Science Foundation Grant No. PHYS-1066293 and the hospitality of the Aspen Center for Physics. This project also made use of SDSS data. Funding for the SDSS and SDSS-II has been provided by the Alfred P. Sloan Foundation, the Participating Institutions, the National Science Foundation, the U.S. Department of Energy, the National Aeronautics and Space Administration, the Japanese Monbukagakusho, the Max Planck Society, and the Higher Education Funding Council for England.  The SDSS Web Site is http://www.sdss.org/. 
The SDSS is managed by the Astrophysical Research Consortium for the Participating Institutions. The Participating Institutions are the American Museum of Natural History, Astrophysical Institute Potsdam, University of Basel, University of Cambridge, Case Western Reserve University, University of Chicago, Drexel University, Fermilab, the Institute for Advanced Study, the Japan Participation Group, Johns Hopkins University, the Joint Institute for Nuclear Astrophysics, the Kavli Institute for Particle Astrophysics and Cosmology, the Korean Scientist Group, the Chinese Academy of Sciences (LAMOST), Los Alamos National Laboratory, the Max-Planck-Institute for Astronomy (MPIA), the Max-Planck-Institute for Astrophysics (MPA), New Mexico State University, Ohio State University, University of Pittsburgh, University of Portsmouth, Princeton University, the United States Naval Observatory, and the University of Washington. 

{\it Facilities:}  \facility{Sloan ()} \facility{COS ()}

\end{document}